\documentclass[fleqn,usenatbib]{mnras}
\usepackage{newtxtext,newtxmath}
\usepackage[T1]{fontenc}
\usepackage{ae,aecompl}
\usepackage{graphicx}
\usepackage{amsmath}
\usepackage{amssymb}
\usepackage[dvipsnames]{xcolor}
\usepackage{natbib}
\usepackage[utf8]{inputenc}
\usepackage{multirow}
\usepackage{xtab, afterpage}
\usepackage{pdflscape}
\usepackage{dcolumn}
\usepackage{xstring}
\usepackage{collcell}
\usepackage{array}
\usepackage[stable]{footmisc}
\usepackage{longtable}
\usepackage{booktabs}
\usepackage{float}
\usepackage{enumitem}
\usepackage{bm}
\restylefloat{table}
\usepackage{caption}
\newcommand{\AddPhantomMinusIfNeeded}[1]{%
\IfDecimal{#1}{
\IfBeginWith{#1}{-}{\ensuremath{#1}}{\ensuremath{\phantom{-}#1}}%
}{%
\ensuremath{\phantom{-}}#1
}%
}%
\newcommand{\daa}{\Delta \alpha/\alpha_0}

\newcommand{\fev}{{Fe~V}}

\newcolumntype{L}{>{\collectcell\AddPhantomMinusIfNeeded}{l}<{\endcollectcell}}

\title[Fine structure constant on a white dwarf surface]{Measuring the fine structure constant on a white dwarf surface; a detailed analysis of Fe V absorption in G191-B2B}

\author[J.Hu {\it et al}]{
J. Hu$^{1}$\thanks{Email: jitinghu@swin.edu.au},
J. K. Webb$^{2}$\thanks{Email: jkw@phys.unsw.edu.au},
T. R. Ayres$^{3}$,
M. B. Bainbridge$^{4}$,
J. D. Barrow$^{5}$,\newauthor
M. A. Barstow$^{4}$,
J. C. Berengut$^{2}$,
R. F. Carswell$^{6}$,
V. Dumont$^{7}$,
V. Dzuba$^{2}$,\newauthor
V. V. Flambaum$^{2}$,
C. C. Lee$^{5}$,
N. Reindl$^{8}$,
S. P. Preval$^{4}$,
W.-Ü L. Tchang-Brillet$^{9}$.
\\ \\
$^{1}$ADACS, Swinburne University of Technology, Hawthorn, VIC 3122, Australia\\
$^{2}$School of Physics, University of New South Wales, Sydney, NSW 2052, Australia\\
$^{3}$Center for Astrophysics and Space Astronomy, University of Colorado, 389 UCB, Boulder, Colorado 80309-0389, USA\\
$^{4}$Department of Physics and Astronomy, University of Leicester, University Road, Leicester LEI 7RH, UK\\
$^{5}$DAMTP, Centre for Mathematical Sciences, University of Cambridge, Cambridge CB3 0WA, UK\\
$^{6}$Institute of Astronomy, Madingley Road, Cambridge CB3 0HA, UK\\
$^{7}$Computational Research Division, Lawrence Berkeley National Laboratory, Berkeley, CA 94720, USA\\
$^{8}$Institute for Physics and Astronomy, University of Potsdam, Karl-Liebknecht-Str. 24/25 14476 Potsdam, Germany\\
$^{9}$LERMA, Observatoire de Paris-Meudon, PSL Research University, CNRS UMR8112, Sorbonne Université, F-92195 Meudon, France
}

\date{Received XXX. Accepted YYY.}
\pubyear{2020}

\begin{document}
\label{firstpage}
\pagerange{\pageref{firstpage}--\pageref{lastpage}}
\maketitle

\begin{abstract}
The gravitational potential $\phi = GM/Rc^2$ at the surface of the white dwarf G191-B2B is 10,000 times stronger than that at the Earth's surface. Numerous photospheric absorption features are detected, making this a suitable environment to test theories in which the fundamental constants depend on gravity. We have measured the fine structure constant, $\alpha$, at the white dwarf surface, used a newly calibrated Hubble Space Telescope STIS spectrum of G191-B2B, two new independent sets of laboratory Fe V wavelengths, and new atomic calculations of the sensitivity parameters that quantify Fe V wavelength dependency on $\alpha$. The two results obtained are: $\daa = 6.36 \pm (0.33_{\text{stat}} + 1.94_{\text{sys}}) \times 10^{-5}$ and $\daa = 4.21 \pm  (0.47_{\text{stat}} + 2.35_{\text{sys}}) \times 10^{-5}$. The measurements hint that the fine structure constant increases slightly in the presence of strong gravitational fields. A comprehensive search for systematic errors is summarised, including possible effects from line misidentifications, line blending, stratification of the white dwarf atmosphere, the quadratic Zeeman effect and electric field effects, photospheric velocity flows, long-range wavelength distortions in the HST spectrum, and variations in the relative Fe isotopic abundances. None fully account for the observed deviation but the systematic uncertainties are heavily dominated by laboratory wavelength measurement precision.
\end{abstract}


\section{Introduction}

General relativity has passed all weak-field observational and experimental tests to date. Nevertheless, the theory predicts singularities on small scales and is incompatible with quantum field theory. To be compatible with cosmological observations, the theory also requires most of the energy content of the universe to be in the form of an unknown dark energy. These things lead us to expect that GR will ultimately become the low-energy limit of some future more fundamental unification theory. Searches for departures from standard GR, particularly in stronger-field situations, are important in this respect. This is the aim of the work described in this paper.

Fundamental constants may vary in the presence of strong gravitational fields, a possibility first proposed by \citet{Dicke1959, Dicke1964}, the latter republished in \citet{Dicke2019}, with a related discussion given in \citet{Bekenstein1982}. More recent ideas concerning quantum gravity unification theories allow the possibility of space and time variations in the low-energy `constants' of Nature, either because of the presence of extra space dimensions or the non-uniqueness of the quantum vacuum state for the universe, e.g. \citet{uzan11}. Strong-field tests using pulsar timing observations in a triple stellar system place tight constraints on departures from standard General Relativity \citep{Archibald2018, Voisin2020}.

Since the gravitational potential at the surface of a white dwarf is typically 4 to 5 orders of magnitude stronger than on Earth, white dwarf atmospheres provide an interesting environment to search for new physics or new tests of the weak equivalence principle. Discussions concerning relativistic effects in white dwarfs include \citet{Wheeler1968, Jain2016, Carvalho2018}. If relativistic effects are weak (as is the case for white dwarf surfaces; $GM/Rc^{2}\sim 10^{-4}$, within an order of magnitude of the gravitational potential perturbation observed on the last scattering surface of the microwave background), the total scalar charge is proportional to the number of nucleons in the object \cite{flambaum08}.

Near massive gravitating bodies, different types of couplings between scalar fields and other fields can lead to an increase or decrease in the coupling constant strengths \cite{magueijo02}. These scalar fields can be the carriers of variations in traditional constants of physics, like $G,\alpha$ and $m_{e}/m_{p}$ and so precision studies of white dwarf atmospheres offers a new laboratory for fundamental physics that is not available on Earth. For small variations, the scalar field variation is proportional to the change in the dimensionless gravitational potential, $\phi =GM/Rc^{2}$ and hence \emph{to} the compactness ($M/R$) of an object. Compact objects with high mass and small radius, could thus exhibit variations in the fine structure `constant', $\alpha$, for example, with a relative change given by
\begin{equation}
\frac{\Delta \alpha }{\alpha _{0}}=\frac{\alpha _{\phi }-\alpha _{0}}{\alpha
_{0}}\propto \Delta \phi ,  \label{eq:a-g-dependence}
\end{equation}
where $\Delta \phi$ is a change in the gravitational potential (in this case the difference between the white dwarf and terrestrial values). Inhomogeneous variations in the cosmological setting were studied in \cite{Barrow2002, Mota2004, Mota2004b, Barrow2003}.

In this paper, we present a detailed analysis of the spectrum of G191-B2B to measure $\Delta \alpha /\alpha _{0}$. The first such analysis was reported in \citet{berengut13}. Both that paper and this make use of the \textit{Many Multiplet method} \citep{dzuba99b, webb99, dzuba03, flambaum09}. However, a new analysis of G191-B2B is required because: (i) two new samples of Fe V laboratory wavelength measurements have since been published, (ii) the G191-B2B HST STIS spectra themselves have been refined/re-reduced, in principle providing a more accurate calibration, (iii) the \citet{berengut13} analysis used simple line centroid measurements, which do not readily identify line blends or other anomalies. Here we use a different approach -- simultaneously fitting many Fe V transitions using Voigt profiles (which effectively means that we model the data using far fewer positional parameters), and (iv) the coefficients used to parameterise each transition's sensitivity to a change in $\alpha $ have been independently re-calculated (Section \ref{subsec:q}). Having 2 independent sets yields an approximation for associated uncertainties.

The structure of the remainder of this paper is as follows: Section \ref{sec:WDs} provides a brief reminder of basic white dwarf astrophysics and describes previous constraints. Section \ref{sec:Astrodata} describes the HST/STIS spectra of the white dwarf G191-B2B. Section \ref{sec:Atomdata} details all atomic/laboratory data used. Section \ref{sec:definesample} discusses line detection in the G191-B2B spectrum, matching those lines with the laboratory Fe V wavelengths, and the further refinement of the sample to remove weak and asymmetric features. Section \ref{sec:Measureingalpha} presents the modelling procedures for measuring $\Delta \alpha/\alpha_0$ and Section \ref{sec:Uncertainties} summarises the results of numerical calculations quantifying several possible sources of systematic error. 

Sections \ref{sec:Results} and \ref{sec:Discussion} present and discuss the
final results.

\section{Observational constraints using white dwarfs}
\label{sec:WDs}

White dwarfs are the most common end-products of stars. More than 95\% of stars in our Galaxy will end up as white dwarfs. Stars with initial masses of $\sim \!\!$ (1--8) $M_{\odot}$ eventually evolve into white dwarfs \cite[]{fontaine13}. As stars evolve, the energy generated by nuclear interactions fails to balance gravity and collapse occurs. Stable radii are reached when electron degeneracy pressure balances gravity \cite[]{koester90}, leaving very compact, high surface gravity remnants, white dwarfs. A typical white dwarf has an oxygen/carbon core surrounded by a thin hydrogen/helium (H/He) envelope. Due to processes such as levitation \cite[]{chayer95} and debris accretion from disrupted planetary systems, interstellar material, or comets \cite[]{zuckerman03}, metallic elements are found in the atmosphere.

White dwarfs are classified according to spectral features (e.g. atmospheric composition and colour), DA white dwarfs being those with a hydrogen enriched atmosphere \cite[]{sion83}. These white dwarfs are useful test-beds for investigating dependence of fine-structure constant variation on strong gravitational field. Some non-DA white dwarfs that have similar characteristics can also be used, such as the hot sub-dwarf O star, BD+$28^{\circ}4211$. White dwarf characteristics that are desirable for $\alpha$ measurements include (i) 
a strong surface gravitational field; DA white dwarfs can have surface gravity $\log g$ between $\sim$7 and $\sim$8 \cite[]{bergeron92}, and (ii) multiple heavy element transitions; various species, including {\fev} or Ni V have been observed in white dwarf spectra \cite[]{sion92, holberg94, werner94}. These highly ionised lines are particularly useful for measuring variation in fine-structure constant ($\alpha$), as they are very sensitive to a change in $\alpha$.

Hot DA white dwarfs with metal lines typically have effective temperatures $T_{\text{eff}} \geq 50,000$ K \cite[]{Barstow2003, Barstow2014}. The {\it Space Telescope Imaging Spectrograph (STIS)} provides the highest resolution available for UV spectroscopy, the E140H grating having a resolving power of 114,000. Detailed analyses of the white dwarf G191-B2B are reported in \citet{preval13,Rauch2013}. By comparing the observed data with NLTE model spectra, many Fe V and Ni V absorption features were identified.

The first astronomical measurement of $\alpha$ variation in strong gravitational fields \citep{berengut13} used the \citet{preval13} spectrum and line identifications. Applying the many-multiplet method \citep{dzuba99b, webb99}, two $\daa$ measurements using {\fev} and Ni V lines separately gave $\daa$({\fev}) $= (4.2 \pm 1.6) \times 10^{-5}$, and $\daa$(Ni V) $= (-6.1 \pm 5.8) \times 10^{-5}$. The discrepancy between these two measurements suggested significant laboratory wavelength systematics in either the {\fev} or the Ni V line lists, or both.

Analyses of additional spectra are described in \cite{Bainbridge2017a, Bainbridge2017b}, who discuss the possibility of making ${\daa}$ measurements over a wide range in surface gravity using 8 white dwarfs, including G191-B2B. Collectively, the 8 results based on {\fev} hint at a non-zero (positive) ${\daa}$ and a possible correlation with field strength. However, since those measurements are highly preliminary and include no estimate of systematics, we defer any further quantitative comments regarding those spectra to a later paper.

\section{Astronomical data}
\label{sec:Astrodata}

The analysis described in this paper is based on a high signal to noise STIS FUV spectrum of the white dwarf G191-B2B. The observations, data reduction, and other details are described fully in \citet{hu19}, so only a brief summary is given here. G191-B2B is a common calibration source for STIS so the total integration time is high. In total, 37 archival E140H exposures of G191-B2B using the $0.2\times 0.2$'' aperture were used. The E140H observations were conducted between 1998 and 2009 in observing cycles 7, 8, 10, and 17. The total E140H exposure time is 17.36 hours. In addition to the E140H exposures, which cover the wavelength range (1150--1700) {\AA}. There are also 5 exposures taken between 2000 and 2009 in observing cycles 9, 10 and 17, using the shortest wavelength setting (1763 {\AA}) with the E230H grating covering the range (1630--1897) {\AA}. The full coverage of the spectrum spans 1150 {\AA} to 1897 {\AA}. The spectral resolution is $\lambda/\Delta\lambda \approx 114,000$.

\section{Atomic data} \label{sec:Atomdata}

\subsection{Laboratory \& Ritz wavelengths} \label{subsec:atomic-wav}

The analysis of G191-B2B by \cite{berengut13} showed that wavelength errors dominate the $\daa$ uncertainty. That analysis used the best available wavelengths at that time \citep{ekberg75}. Two more new Fe V wavelength datasets have been published since then: the Ritz calculations of \cite{kramida14}, and the laboratory measurements by \cite{ward19} at NIST. For the measurement of $\daa$ presented in this paper, we used the Fe V electric dipole (E1) transitions available from these two new wavelength datasets (the majority being $3d^34s$-$3d^34p$ transitions) that lie in the range 1149 {\AA} to 1705 {\AA}. We now comment briefly on each set of Fe V laboratory wavelengths used in our analysis.\\

\noindent{\it Kramida 2014} (hereafter, K14): \\

This set of Fe V lines is from \citet{kramida14}. He used previous laboratory measurements of around 2000 wavelengths, including \citet{ekberg75}, \citet{azarov01}, \citet{fawcett74}, \citet{kalinin85}, with the lines in the wavelength range of interest to this study being dominated by the measurements of \citet{ekberg75}. \citet{kramida14} applied a least squares fitting method \cite[]{Kramida11} to determine optimised energy levels and hence Ritz wavelengths. Since multiple spectral lines are used to determine the energy levels, Ritz wavelengths generally have smaller uncertainties than individual laboratory experiment sets. However, this also means that wavelength calibration errors affect the Ritz wavelengths in a complex way as they are determined from spectral lines in different wavelength regions. Within our wavelength range of interest, the uncertainties lie in the range is 1.5 to 8  m{\AA}.\\

\noindent{\it Ward {\it et al} 2019} (hereafter, W19): \\ 
This set of {\fev} lines comes from an experiment to measure {\fev} and other species at the National Institute of Standards and Technology, Gaithersburg, Maryland USA \cite[]{ward15, ward15a, ward19}. The experiment measured 164 Fe V lines between 1200 {\AA} and 1600 {\AA}. In this sample, uncertainties are assigned to each Fe V wavelength and lie in the range 2.4 to 10 m{\AA}.\\

We opt to remove the least accurate Fe V wavelengths from both samples. For K14 we discard all lines with uncertainties greater than 4 m{\AA}. The 4 m{\AA} cut is mildly arbitrary but was chosen because it is the measurement uncertainty for the Fe V sample from \cite{ekberg75} on which the optimised K14 sample is based and in order retain a significant number lines in the sample. This leaves 377 lines. For W19, we do the same (but see the caption to Table \ref{tab:allwavelengths}). This leaves 129 lines.

\begin{figure*}
\includegraphics[width=12cm]{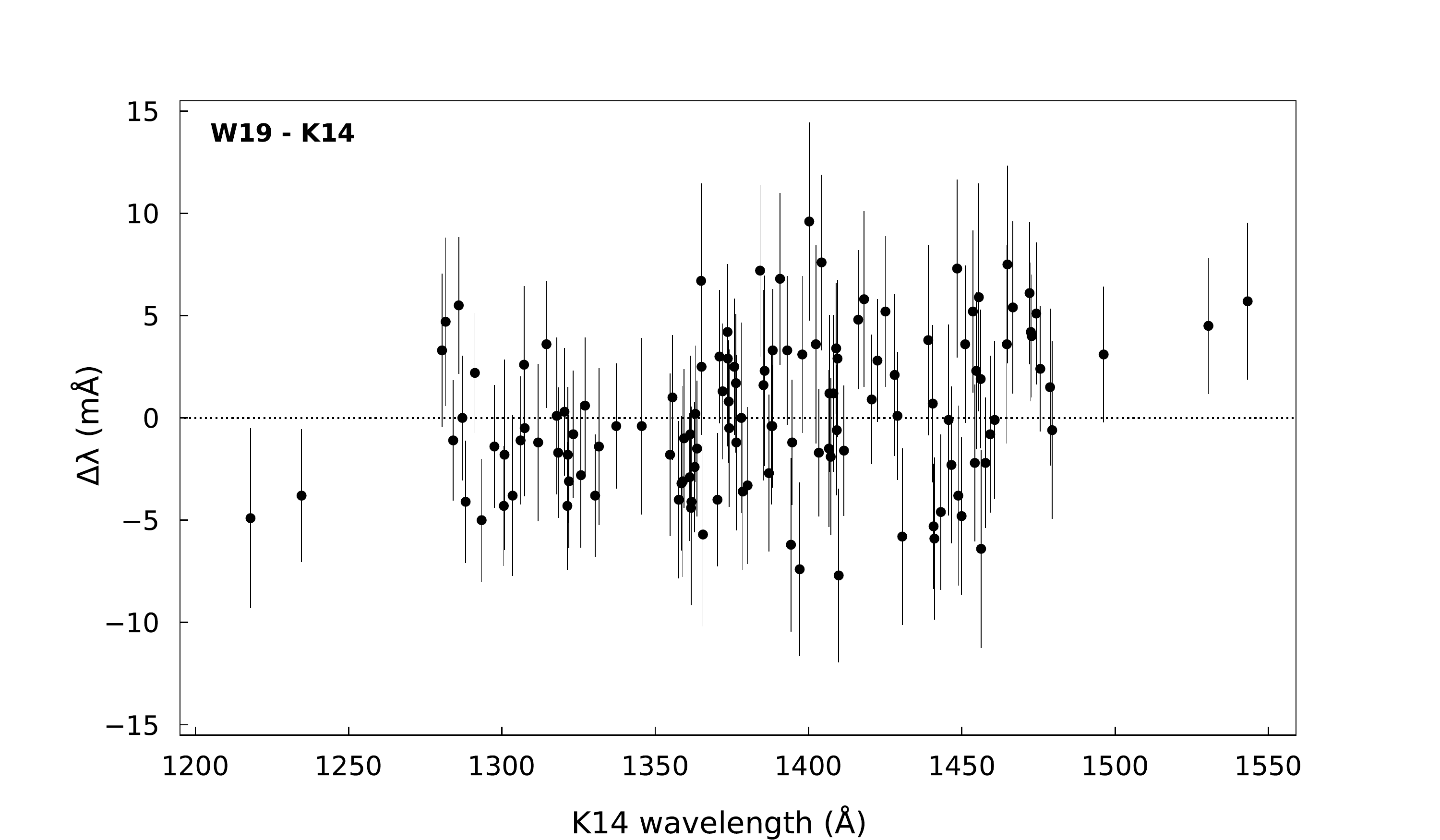}\centering
\caption{The difference between the W19 and K14 wavelengths. See \citet{ward19} for a detailed comparison between Fe~V wavelength sets.}
\label{fig:wav-comparison01}
\end{figure*}

To examine consistency, in Figure \ref{fig:wav-comparison01} we plot wavelength difference for pairs of lines from the two wavelength datasets. $\Delta \lambda$, the difference between W19 and K14 wavelength, has a mean value of 0.2 m{\AA}, with a standard deviation of 3.7 m{\AA}.

\subsection{Other atomic data}

Since we are fitting Voigt profiles to the observed Fe V absorption lines in the G191-B2B atmosphere, the column density parameter returned from the Voigt profile fitting is only a measure of line strength (and does not give a true column density estimate). To calculate Voigt profile models, we necessarily require values for the oscillator strength $f$ and damping constant $\Gamma$. The $f$ values are taken from \citet{kramida14}, supplemented by 18 missing values from the Kurucz database\footnote{\url{http://kurucz.harvard.edu/}}. $\Gamma$ values are taken from \citet{aggarwal17} where available and Kurucz’s values used otherwise. However, in our modelling procedure, each observed transition is assigned its own column density and velocity dispersion parameter ($b$-parameter). Only relative line positions are tied in the fitting procedure. This means that the oscillator strengths and damping coefficients are effectively treated as free parameters.

\subsection{Sensitivity coefficient calculations} \label{subsec:q}

The wavenumber $\omega$ of a transition measured in the white dwarf rest-frame is given by
\begin{equation} \label{eq:omega}
\omega = \omega_0 + qx,
\end{equation}
where $\omega_0$ is the laboratory wavenumber, $q$ is the sensitivity coefficient to changes in $\alpha$, and $x=(\alpha_{\phi}/\alpha_0)^2 -1$ \citep{dzuba99, webb99}.

New {\fev} $q$-coefficients for transitions involving $3d^34s$ and $3d^34p$ configurations have been calculated. The same approach as in \citet{ong13} has been followed, but the new calculations include about a hundred new states, leading to an enormous number of new electric dipole transitions. The calculations use the combined configuration interaction and the many body perturbation theory method, CI+MBPT \citep{dzuba96,dzuba98}. The B-spline technique \citep{johnson86} is used to construct a single-electron basis set. The calculations are done in the $V^{N-4}$ approximation \citep{dzuba05}, which means that the initial self-consistent Hartree-Fock procedure is done for the closed-shell Ar-like Fe~IX ion with all four valence electrons removed. The basis states in valence space are calculated in the frozen field of the Ar-like Fe~IX core. The CI technique is used to construct four-electron valence states while MBPT is used to include core-valence correlations (see \cite{ong13,dzuba96,dzuba98} for further details).

An analysis of the accuracy of the calculations is described in \cite{ong13}. This includes calculations with different initial approximations ($V^N$, $V^{N-1}$, etc.) which give very close results. Therefore, in the present paper we use just one approach based on the $V^{N-4}$ potential. It is slightly different from the methods used in \cite{ong13} and independent computer codes were used. Comparing the two sets of coefficients therefore provides an empirical estimate of the $q$-coefficient uncertainties. In the end, we believe that the $q$-coefficient uncertainty does not exceed 10\%. For Fe V lines to be used in measuring $\daa$, we must of course have $q$-coefficients; 345 of the 377 in the K14 sample and 120 of the 129 in the W19 sample have these.

Inspection of Table \ref{tab:allwavelengths} shows that some transitions have not been assigned $q$-coefficients. The reason for this is  that due to the high density of excited states, matching the observed and calculated energy levels is sometimes ambiguous. This means we cannot reliably assign $q$-coefficients to some of the observed transitions. Therefore, we have not presented values of $q$ in some cases. We noted (subsequent to the completion of this work) that matching energy levels can be clarified using the procedures of \cite{kramida13}. This is a possible area to be developed in future work so that more transitions can be used.

\begin{table*}[!ht]
\centering
\resizebox{\textwidth}{!}
{
\bgroup
\def\arraystretch{1.1}
\setlength\tabcolsep{3pt}
\begin{tabular}{r*{11}{l}@{\extracolsep{2pt}}*{5}{l}@{}*{1}{r}}
\hline
\multicolumn{1}{c }{\multirow{2}{*}{\textbf{ID}}} &
\multicolumn{1}{c }{\multirow{2}{*}{\textbf{C1}}} &
\multicolumn{1}{c }{\multirow{2}{*}{\textbf{T1}}} &
\multicolumn{1}{c }{\multirow{2}{*}{\textbf{J1}}} &
\multicolumn{1}{c }{\multirow{2}{*}{\textbf{C2}}} &
\multicolumn{1}{c }{\multirow{2}{*}{\textbf{T2}}} &
\multicolumn{1}{c }{\multirow{2}{*}{\textbf{J2}}} &
\multicolumn{1}{c }{\multirow{2}{*}{\textbf{E1 (cm$^{-1}$)}}} &
\multicolumn{1}{c }{\multirow{2}{*}{\textbf{E2 (cm$^{-1}$)}}} &
\multicolumn{3}{c }{\textbf{K14}} &
\multicolumn{3}{c }{\textbf{W19}} &
\multicolumn{1}{c}{\multirow{2}{*}{\textbf{f}}} &
\multicolumn{1}{c}{\multirow{2}{*}{$\bm{\Gamma}$
\textbf{ (s$\bm{^{-1}}$)}}} &
\multicolumn{1}{c}{\multirow{2}{*}{\textbf{$\bm{q}$ (cm$\bm{^{-1}}$)}}} \\ \cline{10-12} \cline{13-15} \cline{13-14}\noalign{\vskip\arrayrulewidth}
&&&&&&&&&
\multicolumn{1}{c }{\textbf{$\bm{\lambda}$ ({\AA})}} &
\multicolumn{1}{c }{} &
\multicolumn{1}{c }{\textbf{$\bm{\delta\lambda}$ ({\AA})}} &
\multicolumn{1}{c }{\textbf{$\bm{\lambda}$ ({\AA})}} &
\multicolumn{1}{c }{} &
\multicolumn{1}{c }{\textbf{$\bm{\delta\lambda}$ ({\AA})}} &
&&\\
\hline
60& 3d3.(2D1).4s & 3D &1& 3d3.(2D1).4p & 3P* &1&258769.6&335267.9&1307.2190&&0.0030&1307.2216&&0.0024&9.6e-02&4.17E+09& 2862  \\
61& 3d3.(2P).4s & 1P &1& 3d3.(2P).4p & 1P* &1&219486.7&295973.0&1307.4242&$\star$&0.0023&1307.4237&$\diamond$&0.0024&2.2e-01&6.80E+09& 2311  \\
62& 3d3.(2G).4s & 3G &4& 3d3.(2P).4p & 3D* &3&209109.9&285474.0&1309.5147&&0.0020&&&&1.1e-02&5.65E+09& 2992  \\
63& 3d3.(2P).4s & 3P &1& 3d3.(2D2).4p & 3P* &0&214611.3&290903.4&1310.7510&&0.0030&&&&6.0e-04&8.06E+09& 3286  \\
64& 3d3.(4P).4s & 3P &2& 3d3.(2D2).4p & 3D* &3&213649.0&289912.8&1311.2378&&0.0019&1311.2293&&0.0074&3.8e-02&5.88E+09& 2389  \\
65& 3d3.(2P).4s & 1P &1& 3d3.(2D2).4p & 1D* &2&219486.7&295716.2&1311.8290&$\star$&0.0030&1311.8278&$\diamond$&0.0024&1.9e-01&1.01E+10& 1616  \\
66& 3d3.(4P).4s & 3P &0& 3d3.(2D2).4p & 3D* &1&212542.0&288669.5&1313.5851&$\star$&0.0023&&&&6.6e-02&6.17E+09& 2507  \\
67& 3d3.(2G).4s & 3G &3& 3d3.(2P).4p & 3D* &2&208838.0&284911.0&1314.5256&$\star$&0.0017&1314.5292&$\diamond$&0.0026&2.2e-03&5.21E+09& 2920  \\
68& 3d3.(2D1).4s & 3D &2& 3d3.(2D1).4p & 3P* &2&258628.6&334509.2&1317.8610&&0.0030&1317.8611&&0.0024&7.0e-02&4.20E+09&   \\
69& 3d3.(2G).4s & 3G &3& 3d3.(2H).4p & 3H* &4&208838.0&284690.3&1318.3505&&0.0021&1318.3488&&0.0024&6.0e-02&5.99E+09& 3660  \\
\hline
\end{tabular}
\egroup
}
\vspace{0.1in}
\caption{\label{tab:example_atomic_data}Summary of atomic data. K14 are Ritz wavelengths \citep{kramida14}. Lines that we used for measuring $\daa$ are marked with a $\star$. The bulk of the analysis described in this paper was carried out on a pre-publication version of the W19 tabulated FeV lines that had been assigned different uncertainty estimates. The initial cut (accepting only lines with uncertainties $\leq 4$ m{\AA}), and the subsequent filtering, results in our final W19 sample being those lines marked in this table with a $\diamond$. The pre-publication FeV tabulation has not been given here -- in this table we list the W19 data as published in \citet{ward19}. Columns 2-4 and 5-7 are the lower and upper level configuration, term, and J values, adopted from Table 1 in \citet{kramida14} and from NIST (\url{https://physics.nist.gov/PhysRefData/ASD/lines_form.html}). Columns 8 and 9 are the lower and upper energy levels. Oscillator strength $f$ are from \citet{kramida14} if available. 18 lines with missing values are supplemented by data from the Kurucz database ($\dagger$). $\Gamma$ values are taken from \citet{aggarwal17} where available. 15 lines with missing values are supplemented by data from Kurucz database (marked with a $\mathsection$). The $q$-coefficients given in this table are new calculations and supersede those of \citet{ong13}. This table only shows 10 entries. The full table can be found in Table \ref{tab:allwavelengths} in Appendix \ref{appendix:lams_and_qs}.}
\end{table*}

\normalsize

\section{Line detection, identification, and sample refinement} \label{sec:definesample}

\subsection{5$\sigma$ detections using RDGEN}

RDGEN is a multi-purpose program for spectroscopic data analysis \cite[]{rdgen}. One useful function is the detection of absorption lines above a given statistical significance, relative to a user-supplied continuum level. A reliable spectral variance array is required. The algorithm uses the known spectral resolution to constrain the minimum separation between adjacent or blended features. It returns line centroids, equivalent widths, and other relevant measurements, with associated error estimates. Our first step is to measure lines in the G191-B2B spectrum above a 5$\sigma$ detection threshold.

\subsection{Matching the detected and laboratory wavelengths}

The laboratory wavelength sets are derived from terrestrial experiments for which ${\daa}\!=\!0$. The number of absorption lines per unit wavelength interval in the spectrum of G191-B2B is reasonably high, so it may be possible to wrongly associate observed and laboratory lines if the tolerance is insufficiently stringent and/or if in fact ${\daa}\!\ne\!0$ near a white dwarf. If ${\daa}\!\ne\!0$ near a white dwarf, making the assumption that ${\daa}\!=\!0$ may result in line misidentifications (and bias a ${\daa}$ measurement towards zero). Since we do not know {\it a priori} the value of ${\daa}$ at the white dwarf atmosphere, it is therefore preferable to avoid making any assumption about ${\daa}$ when identifying the observed G191-B2B absorption lines. To avoid this problem, line identification is carried out as a function of ${\daa}$, over the range $-10^{-3}\!\!<\!\Delta\alpha/\alpha\!<\!\!10^{-3}$, in steps of $10^{-5}$. The K14 wavelengths have the largest number of lines available (345 lines) so this dataset was used for line identification.

The G191-B2B redshift used is obtained from the measured value of $v=23.8 \pm 0.03$ km s$^{-1}$ \citep{preval13},  obtained using the weighted average of all the identified atmospheric lines (i.e. all identified species, not just {\fev}).

Re-arranging Eq.\ref{eq:omega} and using $\omega = \omega'(1+z)$, the observed-frame laboratory wavelength becomes
\begin{align} \label{eq:lambda}
\lambda' & = \frac{\lambda_0 (1+z)}{1+qx/\omega_0} \\
& \approx \lambda_0 (1+z) \left(1-\frac{Q\Delta\alpha}{\alpha_0}\right),
\end{align}
where $\omega'$ is the observed-frame wavenumber, $z$ is its redshift (the summed effects of stellar peculiar velocity and gravitational redshift), $\lambda_0$ is the terrestrial laboratory wavelength, the dimensionless quantity $Q\!=\!2q/\omega_0,\, x\!\approx\!2\Delta\alpha/\alpha_0$, and where the first two Taylor series terms have been used to approximate the factor of $1\!/\!(1\!+\!qx\!/\omega_0)$ in Eq.\ref{eq:lambda}.

For a line to be identified as \fev, we require the observed and laboratory wavelengths to agree within 
\begin{equation}
\Delta\lambda = |\lambda_{\text{obs}} - \lambda'| \,\, \lesssim \,\, n \sqrt{\sigma(\lambda_{\text{obs}})^2 + \sigma(\lambda')^2},
\label{tolerance}
\end{equation}
where $\lambda_{\text{obs}}$ and $\lambda'$ are the observed-frame white dwarf and laboratory wavelengths. The error contribution due to the white dwarf redshift is small compared to the other terms so has been ignored.

\begin{figure*}
\centering
\includegraphics[width=\textwidth]{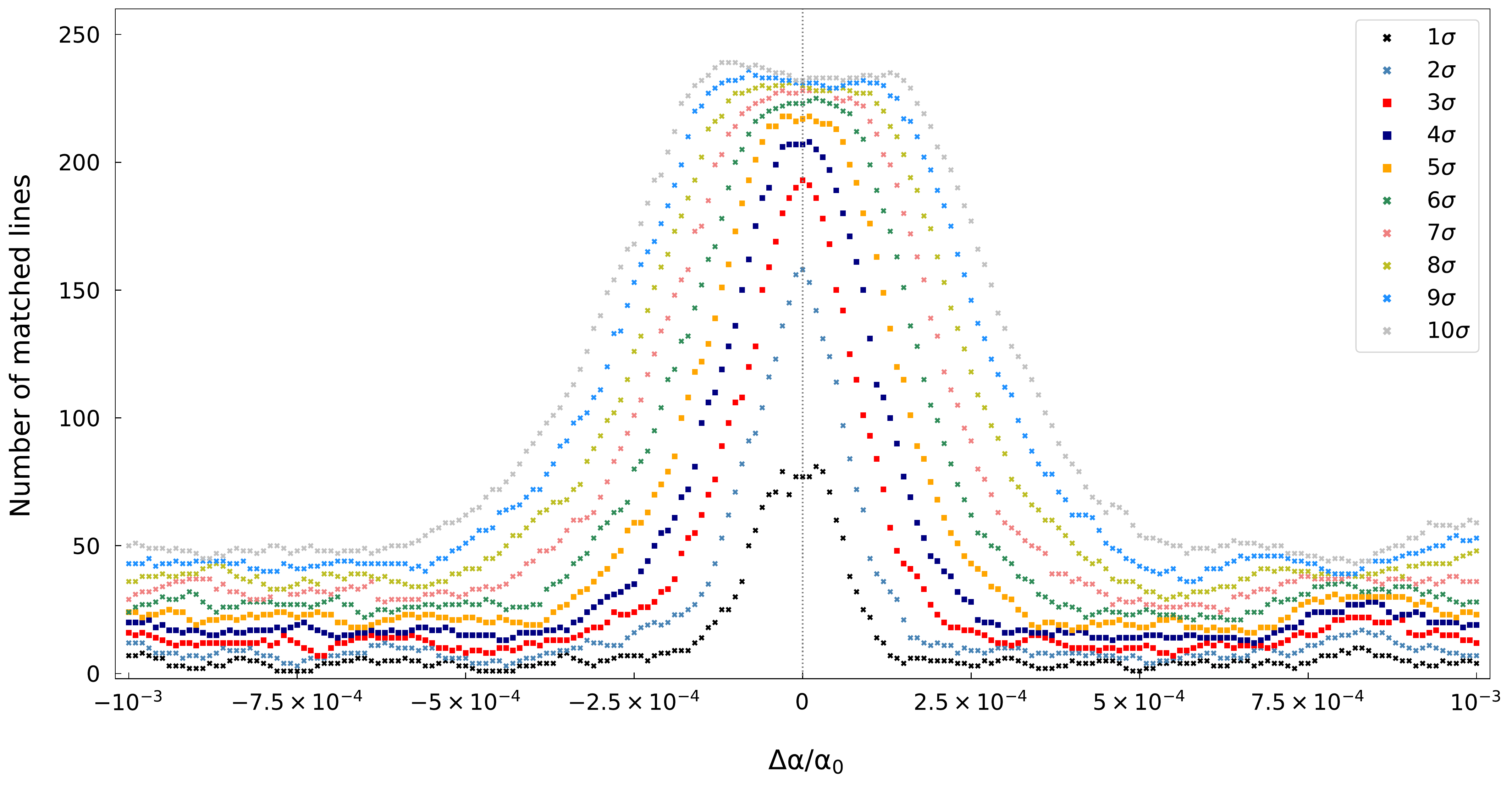}
\caption{This figure illustrates the identification Fe{\sc v} lines in the spectrum of G191-B2B without assuming that $\Delta\alpha/\alpha = 0$, avoiding measurement bias. Lines are identified by stepping through $-10^{-3} < \Delta\alpha/\alpha < 10^{-3}$ in steps of $10^{-5}$, modifying the Fe{\sc v} laboratory wavelengths according to Eq.\ref{eq:lambda}, then matching laboratory and rest-frame wavelengths using the tolerance defined by Eq.\ref{tolerance}. Each coloured curve corresponds to a different tolerance ($n = 1,2,3,..., 10$) for colours blue, green, grey, ..., red. As the identification criterion is relaxed, more lines are identified and the curves become increasingly flat-topped, as expected.}
\label{fig: line-iden-ong}
\end{figure*}

The results of these calculations for a range of $n$ are illustrated in Figure \ref{fig: line-iden-ong}. As can be seen, line identification maximises at or very close to $\daa=0$. The plateauing at large $n$ occurs because the generous identification criterion results in many accepted identifications for each line. The final tolerance used to select lines is $n=3$. This procedure serves two purposes: first, it indicates that if $\daa$ does depart from zero, it is only by a small amount.  Second, it suggests that even if we identify lines using $\daa=0$ with a 3$\sigma$ tolerance, we are unlikely to bias the final result much.

\subsection{Removing blended and weak lines} \label{sec:removeblends}

The total number of {\fev} absorption lines detected in the G191-B2B spectrum using the above selection parameters was 199. We now refine this list by removing blended features. Where RDGEN detects multiple components in an absorption feature in the observed G191-B2B spectrum and where one of the detected components is {\fev} and the other, or others, are not identified as {\fev}, that whole absorption feature is discarded. This reduces the total number identified of {\fev} lines to 164.

Another means of identifying blended profiles comes from the laboratory measurements. The K14 Fe V list identifies 8 cases where a line is resolved into 2 components on the basis of energy level calculations. These 8 lines are discarded, reducing the total number of identified {\fev} lines to 148.

All remaining lines are fitted with Voigt profiles (convolved with the appropriate line spread function - see Section \ref{sec:Measureingalpha} for details). All line parameters are initially untied, i.e. each line is fitted with 3 parameters, column density, velocity dispersion parameter $b$, and redshift. $\chi^2$ for each fit provides a check on goodness of fit. In a few cases, a high value of $\chi^2$ indicates a non-Voigt line shape that had not been picked up earlier. Therefore these lines are removed from the analysis, discarding all lines that returned a $\chi_n^2 > 1.5$. This cut reduces the number of lines used from the G191-B2B spectrum from 148 to 135 lines.

Although lines have been detected above 5$\sigma$, some of the detected features are weak and do not contribute significantly to the $\daa$ measurement. For this reason we apply a final cut: lines with observed equivalent widths less than 0.002 {\AA} (38 lines) are removed. This step reduces the number of lines used from the G191-B2B spectrum from 135 to 97.

Finally, out of the 97 observed lines, all of which have K14 wavelengths, 63 also have W19 wavelengths.

\begin{figure*}
\centering
\includegraphics[width=15cm]{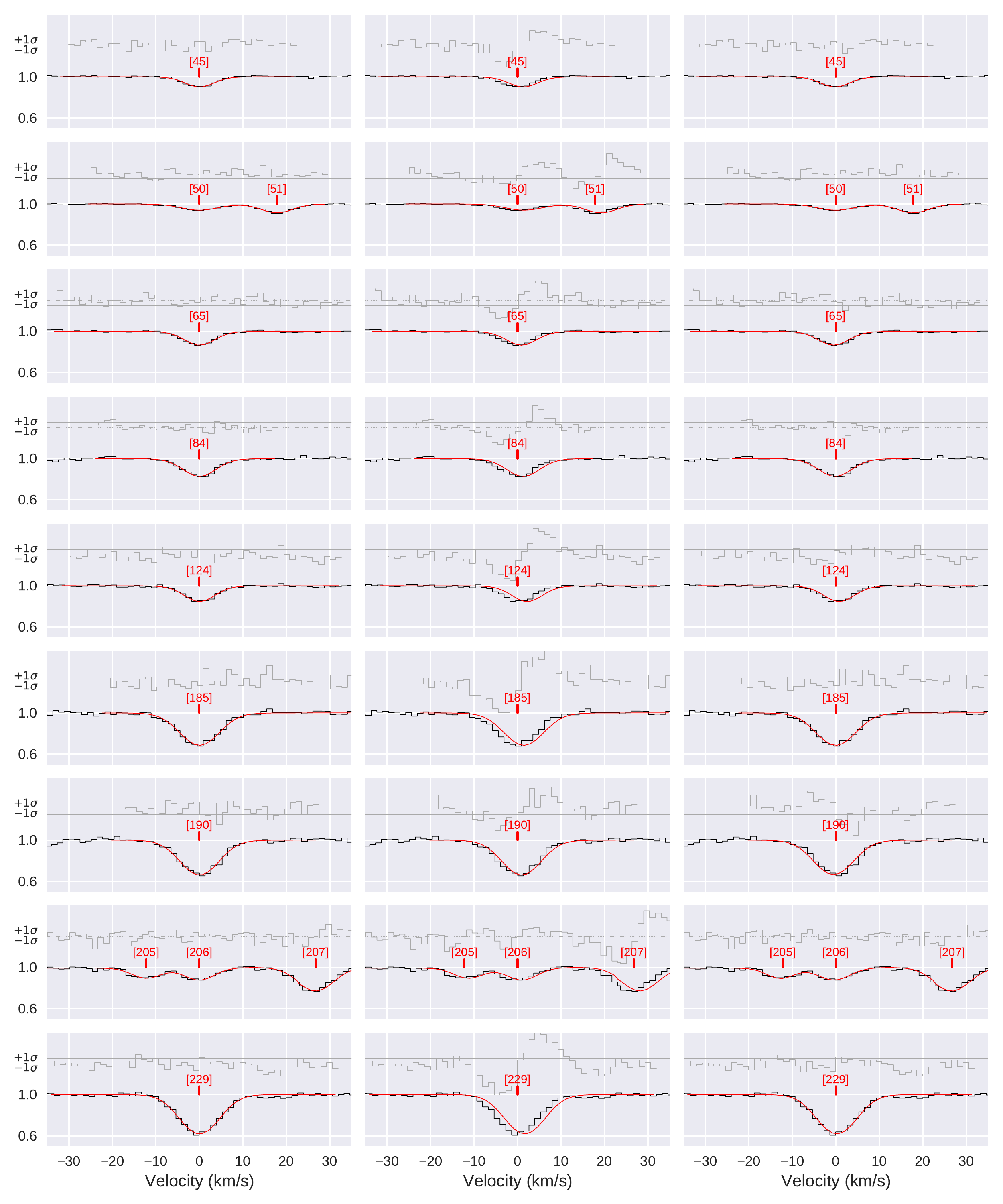}
\caption{Examples of fits to Fe V lines in the G191-B2B E140 STIS spectrum using VPFIT. The $x$-axis is wavelength in {\AA} and the $y$-axis is the normalised flux. The histogram is the data and the Voigt profile fits are over-plotted in red. The ticks indicate the line centres with wavelengths and line IDs (see Table \ref{tab:allwavelengths}. Normalised residuals are illustrated above each profile, the horizontal continuous lines indicate 1$\sigma$ errors. The left hand column shows the best-fit model with $\daa$ as a free parameter. The middle column illustrates the exact same model as the left hand column but with $\daa$ reset to zero. The right hand column shows the best-fit model obtained when $\daa$ is fixed at zero.}
\label{fig:Examplefits}
\end{figure*}

\begin{table*}[!ht]
\centering
\resizebox{0.8\textwidth}{!}
{
\bgroup
\def\arraystretch{1.1}
\setlength\tabcolsep{3pt}
\begin{tabular}{*{5}{r}*{8}{c}}

\hline
\multicolumn{1}{c }{\textbf{ID}} &
\multicolumn{1}{c }{\textbf{$\bm{\lambda_\text{obs}}$}} &
\multicolumn{1}{c }{\textbf{$\bm{\sigma(\lambda_\text{obs})}$}} &
\multicolumn{1}{c }{\textbf{EW}} &
\multicolumn{1}{c }{\textbf{$\bm{\sigma}$(EW)}} &
\multicolumn{1}{c }{\textbf{z [$\bm{\times10^{-5}}$]}} &
\multicolumn{1}{c }{\textbf{$\bm{\sigma}$(z) [$\bm{\times10^{-5}}$]}} &
\multicolumn{1}{c }{\textbf{b}} &
\multicolumn{1}{c }{\textbf{$\bm{\sigma}$(b)}} &
\multicolumn{1}{c }{\textbf{logN}} &
\multicolumn{1}{c }{\textbf{$\bm{\sigma}$(logN)}} &
\multicolumn{1}{c}{\textbf{K14}} &
\multicolumn{1}{c}{\textbf{W19}} \\
\hline

27&1250.833&0.001&0.003&&7.88&0.06&4.99&0.30&12.12&0.02&$\checkmark$& \\
36&1280.573&0.0012&0.00617&0.000197&8.17&0.04&5.36&0.18&13.22&0.01&$\checkmark$&$\checkmark$ \\
38&1281.468&0.001&0.002&&7.93&0.11&5.11&0.59&12.15&0.04&$\checkmark$& \\
43&1286.015&0.001&0.002&&7.73&0.06&3.68&0.30&12.51&0.02&$\checkmark$&$\checkmark$ \\
45&1288.269&0.0009&0.00355&0.000148&7.89&0.04&3.98&0.20&12.73&0.01&$\checkmark$&$\checkmark$ \\
48&1291.298&0.002&0.005&&8.24&0.07&6.50&0.30&12.90&0.02&$\checkmark$&$\checkmark$ \\
50&1293.409&0.001&0.003&&7.85&0.06&5.47&0.27&13.09&0.02&$\checkmark$& \\
51&1293.485&0.001&0.003&&7.92&0.04&4.40&0.18&12.85&0.01&$\checkmark$& \\
52&1297.651&0.001&0.005&&7.95&0.04&5.13&0.17&12.91&0.01&$\checkmark$&$\checkmark$ \\
53&1300.715&0.001&0.0047&0.000162&8.09&0.04&4.59&0.18&12.90&0.01&$\checkmark$& \\
\hline
\end{tabular}
\egroup
}

\vspace{0.1in}
\caption{\label{tab:example_obs_wav} Observed line wavelengths in the STIS spectrum of G191-B2B. ID refers to the number in column 1 of Table \ref{tab:allwavelengths}. $\lambda_\text{obs}$ is the mean observed-frame wavelength of the absorption feature with uncertainty $\sigma(\lambda_\text{obs})$. EW is the equivalent width with uncertainty $\sigma$(EW). The absence of $\sigma$(EW) indicates that another line is nearby such that the detection algorithm interpreted the whole feature as a blend. Wavelength and equivalent widths are in {\AA}. Columns 6-11 give redshift z, b-parameter (in km/s) and column density logN (N is in atoms/cm) with their associated uncertainties. A tick in the last two columns indicates that a line was used for measuring $\alpha$. This table only shows 10 entries. The full table can be found in Table \ref{tab:stiswav} in Appendix \ref{appendix:lams_and_qs}.}
\end{table*}

\section{Measuring $\daa$} \label{sec:Measureingalpha}

To solve for $\daa$, all {\fev} redshift parameters are tied so that the whole sample is parameterised using one single redshift parameter. This removes 96 redshift parameters compared to the modelling described in Section \ref{sec:removeblends}. With all line redshifts tied, VPFIT is then used to solve for $\daa$. The default VPFIT set-up parameters \citep{VPFIT_rfcdoc, VPFIT_overleaf} are used with one important exception. The line spread function (lsf) of the spectrograph and grating (STIS with 140H) is non-Gaussian. Voigt profiles must be convolved with the appropriate lsf\footnote{\url{http://www.stsci.edu/hst/instrumentation/stis/performance/spectral-resolution}} during the modelling process. The non-default VPFIT set-up parameter is the number of sub-bins per pixel, which is set to 13, the number of pixels in the STIS lsf. The free parameters in the fit are thus {\it all} velocity dispersion parameters (one for each absorption line), {\it all} line strengths (``effective column densities'', one for each absorption line), one redshift parameter $z$ for the whole sample, and one $\daa$ parameter.

Once all redshift parameters are tied and the spectrum re-fitted, a small number of previously well-fitted lines produce high $\chi^2$ values. Potential causes of this include:
\begin{enumerate}[leftmargin=*]
\item Line blending that previously went undetected, i.e. an {\fev} line may be correctly identified, but some other species is blended with it (the blended profile remaining reasonably symmetric and hence well-fitted by a single Voigt profile in Section \ref{sec:removeblends}) causing a small shift in the observed wavelength. These would only show up once tied redshifts were introduced;
\item Wavelength calibration errors in the G191-B2B spectrum, although these would necessarily have to be localised, otherwise correlated groups of {\fev} lines would appear discrepant and this seems not to be the case;
\item Uncertainties in individual Fe V laboratory wavelengths;
\item Incorrect $q$-coefficients could cause discrepant $\chi^2$ values for individual lines (although if $\daa=0$, Eq.\ref{eq:omega} shows these have no effect);
\item It is possible (although unlikely) that misidentified lines remain in the sample, i.e. lines which have up until now been assumed to be {\fev} but in fact are some other species.
\end{enumerate}

Regardless of their origin, these points are clipped in an iterative procedure, removing individual spectral fitting regions having normalised $\chi_n\!>\!2$ one at a time, re-fitting for $\daa$ at each iteration. There are a few groups of lines close to each other, thus in the same fitting region. In those cases, if $\chi_n\!>\!2$ for the region as a whole, that region is discarded.

After all selection procedures, the numbers of lines used for the two $\daa$ measurements are 92 with K14 wavelengths and 53 with W19 wavelengths.

\section{Measurement uncertainties} \label{sec:Uncertainties}

Known and potential sources of error have been investigated in considerable detail. Here we briefly described each one and list them in Table \ref{tab:uncertainties}.\\

\noindent{\bf VPFIT measurement uncertainty:}
The VPFIT error (obtained from the covariance matrix at the best-fit solution) depends only on the G191-B2B spectral properties and does not take into account other effects such as laboratory wavelength uncertainties. The method has been checked many times previously, e.g. \citet{King2009}, but even so has been carefully verified again in this context. We generated 1000 synthetic spectra based on the Fe V laboratory wavelengths and the observed G191-B2B spectral properties. Each was fitted and the (known) $\daa$ solved for. The statistical distribution from the 1000 $\daa$ estimates can then be compared to the covariance matrix error estimate. The two values agree well.\\

\noindent{\bf Laboratory wavelength uncertainties:}
In the Fe V laboratory measurements made by \cite{ekberg75}, the nominal line wavelength uncertainties are given as 4 m{\AA}. As noted in \cite{berengut13}, directly comparing the rest-frame G191-B2B and Ekberg wavelengths suggests the actual laboratory uncertainties were somewhat smaller. 

Rather than a single uncertainty estimate, both the K14 and W19 Fe V tabulations provide a range of uncertainties, depending on line quality. For K14, the range for the Fe V lines used to measure $\daa$ here is 1.5 to 4 m{\AA}. For W19, the lines used to measure $\daa$ all have a nominal uncertainty of 3 m{\AA}. 

Additional spectral simulations were therefore also done. First, a model of the white dwarf spectrum is created using the parameters derived from fitting the real spectrum. We perturb each Fe V laboratory wavelength randomly (using the published Fe V wavelength uncertainties) and then re-fit the spectrum using the modified laboratory wavelengths to re-measure $\daa$. The procedure is repeated 10,000 times for each set of laboratory wavelengths (K14 and W19), from which we can empirically determine the scatter in $\daa$ caused by Fe V laboratory wavelength uncertainties. The results of these numerical experiments also show the nominal K14 wavelength uncertainties are over-estimated although the nominal W19 uncertainties are close to correct.\\

\noindent{\bf Pt/Cr-Ne lamp and long-range wavelength distortion:}
The spectrum of G191-B2B used in this analysis is formed by combining 37 different exposures taken over an 11 year period. We do not know how stable the physical parameters of the HST on-board Pt/Cr-Ne emission-line calibration lamp were during time. The wavelength calibration can be also affected by (i) charge transfer inefficiency losses, (ii) instrumental changes such as thermal effects (heating and cooling of the optical bench), (iii) small variations of the position of the target along the spectrograph slit, and (iv) the HST orbital motion. All these factors must combine in some unknown way to produce a net wavelength calibration distortion. 

Since there are no independent calibration sources for comparison, we parameterise any potential distortion using the method described in \citet{dumont17}. Following that procedure, we assume that the effects described above map to a linear distortion in velocity space. An additional free parameter (the velocity distortion slope) is included in the modelling that maps into a small shift on $\daa$ with an additional systematic error term, listed in Table \ref{tab:uncertainties}.\\

\noindent{\bf Multiplet segregation:}
During the modelling procedure, although all line redshifts are tied such that the entire set of Fe V lines is fitted with one single redshift parameter, individual $b$-parameters are free. The reason for this is that there is evidence for stratification in the G191-B2B atmosphere \citep{barstow99, Dreizler1999}, which may imply that a single $b$ value may not necessarily apply to all detected Fe V transitions. This is corroborated by a synthesised model atmosphere\footnote{\url{http://tlusty.oca.eu/}} \citep{tlusty95} for G191-B2B which predicts different atmospheric depths for different Fe V multiplets (also see \citet{Rauch2013}). The computation of synthetic spectra based on TLUSTY calculations are discussed in \citep{Synspec2011}. Spatial segregation, in the presence of a velocity field that varies systematically with atmospheric depth, could in principle emulate varying $\alpha$. Another consequence of Fe V multiplet segregation could be that different multiplets experience different gravitational redshifts. However, both of these effects have been modelled and both turn out to be very small. To illustrate this, using $\Delta z \approx GM\Delta r/r^2 c^2$, assuming a total photospheric height of 10km, the maximum gravitational redshift difference between absorptions created at the top and bottom of the atmosphere is $\Delta z \approx 10^{-8}$. This is an order of magnitude below the {\it statistical} measurement uncertainty on an individual line position, thus very small compared to the net uncertainty on $\daa$.\\

\noindent{\bf Fe isotopic abundances:}
The four stable Fe isotopes have terrestrial relative abundances $^{54}$Fe (5.845\%), $^{56}$Fe (91.754\%), 
$^{57}$Fe (2.119\%), $^{58}$Fe (0.282\%). Deviations from terrestrial abundances in the G191-B2B atmosphere will emulate small (energy level dependent) observed line shifts relative to a model based on terrestrial values. This may emulate a non-zero $\daa$. Variations from terrestrial isotopic abundances have been simulated by appropriately modifying the laboratory wavelengths and re-fitting to the G191-B2B spectrum to examine the impact on $\daa$.\\

\noindent{\bf Zeeman effects:}
If a magnetic field is present, the first order Zeeman effect would broaden or split the observed absorption lines. Two measurements of the magnetic field strength in G191-B2B, derived using independent methods give consistent results: \citet{Bagnulo2018} find $B = -280\pm965$\,G and \citep{hu19} find $B < 2300$\,G (3$\sigma$ upper limit). Using the latter, line shifts due to the quadratic Zeeman effect are $< 10^{-4}$ m{\AA}, four orders of magnitude below Fe V laboratory wavelength uncertainties. The quadratic Zeeman effect therefore has a negligible impact on measuring $\daa$ \citep{hu19}.\\

\noindent{\bf Electric field shifts:}
G191-B2B has a hydrogen rich atmosphere \citep{holberg1991}. Example fundamental stellar parameters are provided by \citet{preval13} and \citet{Barstow2003}: mass 0.52 \(\textup{M}_\odot\), radius 0.0204 \(\textup{R}_\odot\), and surface gravity $GM/R^2 = 3.4 \times 10^5$ m ~s$^{-2}$. Other measurements include those from \citet{Rauch2013} and \citet{Gianninas2011}. At an effective temperature of 52,500K, a TLUSTY atmospheric synthesis model from \citet{preval13} gives a corresponding electron density of $7.0 \times 10^{22}$ m$^{-3}$.

Although we may assume overall electrical neutrality, the proton--electron mass difference gives rise to a gravitational separation between the particles in the atmosphere. This separation in turn produces an opposing electric field. Equilibrium between the forces on protons and electrons gives
\begin{equation}
\label{eq:Electricfield}
E \approx -\frac{m_p}{2e} \frac{GM}{R^2} \approx -1.8 \times 10^{-2}\,\,\textrm{V/m}
 \end{equation}
\citep{alcock1980,koester90}. The quadratic electric field shift for atomic transitions is 
\begin{equation}
\label{EFS}
\Delta \omega \sim \frac{-4\pi\epsilon_0 a_0^3 E^2}{\hbar c}
\end{equation}
\citep{Delone1999} where $\omega$ (=$\nu/c$) is the wavenumber (as in Eq.\ref{eq:omega}), $a_0$ is the Bohr radius, $\epsilon_0$ is the permittivity of free space, and $\hbar$ is the reduced Planck constant.

The corresponding shift $\Delta \omega$ is $\sim 10^{-18}$ cm$^{-1}$. Using Eq. \ref{eq:omega}, approximating $x=2\daa$ so that $\daa = \Delta\omega/q$, and using an illustrative $q=1000$, we obtain $\daa \sim 10^{-21}$. The electric field shift associated with a static electric field is thus negligible.

\begin{table}[!ht]
\centering
\resizebox{\columnwidth}{!}
{
\bgroup
\def\arraystretch{1.1}
\begin{tabular}{r|l|c|c}
\hline 
\hline
\multicolumn{1}{c|}{\textbf{No.}} &
\multicolumn{1}{c}{\textbf{Uncertainty}} &
\multicolumn{1}{|c}{\textbf{$\sigma_{\alpha}$}} &  \multicolumn{1}{|c}{\textbf{Type}} \\
\hline\hline
(1) & Statistical error & 0.33, 0.47 & R \\
\hline
(2) & Laboratory wavelength errors & 1.59(3), 2.00(3) & S \\
\hline
(3) & Long-range distortion & 0.1 & S \\
\hline
(4) & Photosphere stratification - & & \\ & gravitational redshift effects & $\lesssim 0.1$ & S \\
\hline
(5) & Photosphere stratification - & & \\ & bulk velocity flows & $\leq 0.1$ & S \\
\hline
(6) & Fe isotopes - relative & & \\ & abundance variation & $\leq 0.05$ & S \\
\hline
(7) & Zeeman effects & $\leq 10^{-5}$ & S \\
\hline
(8) & Electric field shifts & Negligible
& S \\
\hline
\end{tabular}
\egroup
}
\vspace{0.1in}
\caption{\label{tab:uncertainties}Summary of actual and potential sources of random (R) and systematic (S) uncertainties and their magnitudes or upper limits. $\sigma_{\alpha}$ is the estimated uncertainty on $\daa$ in each case, in units of $10^{-5}$. Where two entries appear, the first number corresponds to the results from the K14 FeV wavelengths, the second from W19.}
\end{table}

\section{Results} \label{sec:Results}

The final fine structure constant measurements at the G191-B2B surface, relative to the terrestrial value, are:
\begin{align}
\daa(K14, {\textrm{92 lines}}) = 6.36 \pm (0.33_{\text{stat}} + 1.94_{\text{sys}}) \times 10^{-5} \\
\daa(W19, {\textrm{53 lines}}) = 4.21 \pm (0.47_{\text{stat}} + 2.35_{\text{sys}}) \times 10^{-5}
\end{align}
The systematic error contributions above are summed over all considered in Table \ref{tab:uncertainties}, including upper limits. The K14 wavelengths produce a result that differs from a null result by 2.8$\sigma$ whilst the W19 wavelengths indicate a 1.5$\sigma$ effect. The error contributions are smaller for the K14 sample because of the larger number of Fe V lines in that sample.

Clearly, the approaches taken in the present paper are very different to the simpler analysis in \citet{berengut13}, which gives $\daa = 4.2 \pm 1.6 \times 10^{-5}$, where the uncertainty estimate is derived only from the nominal wavelength uncertainties of the best available Fe V wavelengths at that time \citep{ekberg75}. In the present analysis we have used more recent independent Fe V laboratory measurements, and have explored and quantified a range of potential systematic errors. Our overall uncertainty remains dominated by laboratory wavelength errors, highlighting the need for more advanced calibration methods in laboratory experiments. Despite this, the agreement between all three results is striking.

\section{Discussion} \label{sec:Discussion}

In this paper, a comprehensive modelling procedure was applied to measure the fine structure constant at the surface of the white dwarf G191-B2B, using a single atomic species, Fe V. The {\it Many-Multiplet method} \citep{dzuba99,webb99} has been used to search for any possible change in $\alpha$ in the presence of a gravitational potential 10,000 times the terrestrial value. 

Wavelengths of Fe V from two laboratory datasets (Kramida 2014; Ward et al. 2019), with improved uncertainties, were compared against the same transitions observed in the G191-B2B photospere. Blended, weak, and asymmetric absorption lines have been identified and removed from the sample to try and minimise associated systematic errors. 

The overall scatter exhibited in Figure \ref{fig:wav-comparison01} is in good agreement with the estimated error bars. Nevertheless, Figure \ref{fig:wav-comparison01} suggests correlated deviations, so some systematic calibration problems may exist in one or both sets of new Fe V wavelengths. In the approximate wavelength range 1290-1360 {\AA }, points lie on average below zero. Whilst the data appears reasonably symmetric about zero above 1350 {\AA }, there is nevertheless a conspicuous clump of high points above $\sim$1470 {\AA }.

To advance this field further, several things are required: (i) it is imperative to improve laboratory wavelength precision by at least one order of magnitude, to $\sim$0.1 m{\AA} or preferably much better, (ii) a large sample of white dwarf spectra should be analysed, and (iii) detailed analyses similar to that described in this paper should be carried out on other species such as FeIV or Ni V.

To summarise, the G191-B2B data calibration/reduction, the Fe V laboratory wavelengths, the $q$-coefficients, and the methodology used in this paper, are all different to those used in \citet{berengut13}. Despite this, the consistency between that earlier result and this more detailed analysis is striking and hints (at the $\sim\!\!3\sigma$ level) that perhaps the fine structure constant increases slightly in the presence of strong gravitational fields.

\section*{Acknowledgements}
We are most grateful to G. Nave, A. Kramida, and J. Ward for extensive comments on a draft of this paper which lead to many improvements. JKW thanks the John Templeton Foundation, the Department of Applied Mathematics and Theoretical Physics and the Institute of Astronomy at Cambridge University for hospitality and support, and Clare Hall for a Visiting Fellowship during this work. JH is grateful for a CSC Fund and a UNSW scholarship. WULTB wishes to acknowledge support from the French CNRS-PNPS national program. MAB, JDB, MBB, SP and NR acknowledge the support of the Leverhulme Trust (Grant number 2015-278).

\section*{Data Availability}

Based on observations made with the NASA/ESA Hubble Space Telescope, obtained from the data archive at the Space Telescope Science Institute. STScI is operated by the Association of Universities for Research in Astronomy, Inc. under NASA contract NAS 5-26555. The Hubble Space Telescope STIS spectra of G191-B2B are freely available in the Barbara A. Mikulski archive.

\newpage

\bibliographystyle{mnras}
\bibliography{paper1} 

\newpage
\onecolumn
\appendix

\section{Atomic data}

\subsection{Fe V wavelengths not used in this analysis} \label{appendix:lams_and_qs}

In addition to the two Fe V laboratory wavelength sets used in this paper, \citet{kramida14} (K14) and \citep{ward19} (W19), there exist line lists from (at least) four previous Fe V laboratory experiments. An early Fe V study is by \citet{bowen37}, who identified 47 $3d^34s$-$3d^34p$ transitions (no quoted uncertainties but 10 m{\AA} implied). \citet{fawcett74} and \citet{kalinin85} measured additional $3d^34s$-$3d^34p$ transitions not included in Ekberg's \citep{ekberg75} line list (measurement uncertainties (10 - 20 m{\AA} and 6 m{\AA} respectively). \citet{azarov01} find $3d^34p$-$3d^35s$ and $3d^34p$-$3d^34d$ transitions (6 m{\AA} uncertainties). These additional sources are noted here for completeness but were not used directly in our work.

\scriptsize
\captionsetup{width=\textwidth}
\setlength\tabcolsep{3pt}
\setlength{\LTcapwidth}{0.8\textwidth}
\def\arraystretch{1.2}
\begin{longtable}{r*{11}{l}@{\extracolsep{2pt}}*{5}{l}@{}*{1}{r}}
\caption{\small{Summary of atomic data. K14 are Ritz wavelengths \citep{kramida14}. Lines that we used for measuring $\daa$ are marked with a $\star$. The bulk of the analysis described in this paper was carried out on a pre-publication version of the W19 tabulated FeV lines that had been assigned different uncertainty estimates. The initial cut (accepting only lines with uncertainties $\leq 4$ m{\AA}), and the subsequent filtering, results in our final W19 sample being those lines marked in this table with a $\diamond$. The pre-publication FeV tabulation has not been given here -- in this table we list the W19 data as published in \citet{ward19}. Columns 2-4 and 5-7 are the lower and upper level configuration, term, and J values for lower levels, adopted from Table 1 in \citet{kramida14} and from NIST (\url{https://physics.nist.gov/PhysRefData/ASD/lines_form.html}). Columns 8 and 9 are the lower and upper level wavenumbers. Oscillator strength $f$ are from \citet{kramida14} if available. 18 lines with missing values are supplemented by data from the Kurucz database ($\dagger$). $\Gamma$ values are taken from \citet{aggarwal17} where available. 15 lines with missing values are supplemented by data from Kurucz database (marked with a $\mathsection$). The $q$-coefficients given in this table are new calculations and supersede those of \citet{ong13}.}}
\label{tab:allwavelengths} \\

\hline
\multicolumn{1}{c }{\multirow{2}{*}{\textbf{ID}}} &
\multicolumn{1}{c }{\multirow{2}{*}{\textbf{C1}}} &
\multicolumn{1}{c }{\multirow{2}{*}{\textbf{T1}}} &
\multicolumn{1}{c }{\multirow{2}{*}{\textbf{J1}}} &
\multicolumn{1}{c }{\multirow{2}{*}{\textbf{C2}}} &
\multicolumn{1}{c }{\multirow{2}{*}{\textbf{T2}}} &
\multicolumn{1}{c }{\multirow{2}{*}{\textbf{J2}}} &
\multicolumn{1}{c }{\multirow{2}{*}{\textbf{E1 (cm$^{-1}$)}}} &
\multicolumn{1}{c }{\multirow{2}{*}{\textbf{E2 (cm$^{-1}$)}}} &
\multicolumn{3}{c }{\textbf{K14}} &
\multicolumn{3}{c }{\textbf{W19}} &
\multicolumn{1}{c}{\multirow{2}{*}{\textbf{f}}} &
\multicolumn{1}{c}{\multirow{2}{*}{$\bm{\Gamma}$
\textbf{ (s$\bm{^{-1}}$)}}} &
\multicolumn{1}{c}{\multirow{2}{*}{\textbf{$\bm{q}$ (cm$\bm{^{-1}}$)}}} \\ \cline{10-12} \cline{13-15} \cline{13-14}\noalign{\vskip\arrayrulewidth}
&&&&&&&&&
\multicolumn{1}{c }{\textbf{$\bm{\lambda}$ ({\AA})}} &
\multicolumn{1}{c }{} &
\multicolumn{1}{c }{\textbf{$\bm{\delta\lambda}$ ({\AA})}} &
\multicolumn{1}{c }{\textbf{$\bm{\lambda}$ ({\AA})}} &
\multicolumn{1}{c }{} &
\multicolumn{1}{c }{\textbf{$\bm{\delta\lambda}$ ({\AA})}} &
&&\\
\hline
\endfirsthead

\caption* {Table \ref{tab:allwavelengths} Continued: Summary of laboratory and Ritz wavelength}\\
\hline
\multicolumn{1}{c }{\multirow{2}{*}{\textbf{ID}}} &
\multicolumn{1}{c }{\multirow{2}{*}{\textbf{C1}}} &
\multicolumn{1}{c }{\multirow{2}{*}{\textbf{T1}}} &
\multicolumn{1}{c }{\multirow{2}{*}{\textbf{J1}}} &
\multicolumn{1}{c }{\multirow{2}{*}{\textbf{C2}}} &
\multicolumn{1}{c }{\multirow{2}{*}{\textbf{T2}}} &
\multicolumn{1}{c }{\multirow{2}{*}{\textbf{J2}}} &
\multicolumn{1}{c }{\multirow{2}{*}{\textbf{E1 (cm$^{-1}$)}}} &
\multicolumn{1}{c }{\multirow{2}{*}{\textbf{E2 (cm$^{-1}$)}}} &
\multicolumn{3}{c }{\textbf{K14}} &
\multicolumn{3}{c }{\textbf{W19}} &
\multicolumn{1}{c}{\multirow{2}{*}{\textbf{f}}} &
\multicolumn{1}{c}{\multirow{2}{*}{\textbf{$\bm{\Gamma}$( s$\bm{^{-1}}$)}}} &
\multicolumn{1}{c}{\multirow{2}{*}{\textbf{$\bm{q}$  (cm$\bm{^{-1}}$)}}} \\ \cline{10-12} \cline{13-15}
&&&&&&&&&
\multicolumn{1}{c }{\textbf{$\bm{\lambda}$ ({\AA})}} &
\multicolumn{1}{c }{} &
\multicolumn{1}{c }{\textbf{$\bm{\delta\lambda}$ ({\AA})}} &
\multicolumn{1}{c }{\textbf{$\bm{\lambda}$ ({\AA})}} &
\multicolumn{1}{c }{} &
\multicolumn{1}{c }{\textbf{$\bm{\delta\lambda}$ ({\AA})}} &
&&\\
\hline
\endhead
\hline
\endfoot
\hline
\endlastfoot
0& 3d3.(2D2).4s & 1D &2& 3d3.(2F).4p & 1D* &2&220620.8&307644.0&1149.1179&&0.0020&&&&2.7e-02&4.78E+09& 2143  \\
1& 3d3.(2F).4p & 3G* &5& 3d3.(2H).4d & 3G &5&307064.1&393894.9&1151.6660&&0.0040&&&&4.6e-02&8.32E+09$^\mathsection$&   \\
2& 3d3.(2H).4p & 3G* &4& 3d3.(2G).4d & 3G &5&292430.6&379110.4&1153.6720&&0.0030&&&&1.2e-02&1.23E+10$^\mathsection$&   \\
3& 3d3.(2G).4p & 3G* &5& 3d3.(4F).4d & 3F &4&280039.5&366564.3&1155.7380&&0.0040&&&&2.4e-02&1.18E+10$^\mathsection$&   \\
4& 3d3.(2F).4p & 3D* &1& 3d3.(2H).4d & 3F &2&308671.3&395049.7&1157.6970&&0.0040&&&&6.0e-02&1.15E+10$^\mathsection$&   \\
5& 3d3.(2H).4p & 3G* &4& 3d3.(2G).4d & 3G &4&292430.6&378763.2&1158.3120&&0.0030&&&&3.7e-03$^\dagger$&1.23E+10$^\mathsection$&   \\
6& 3d3.(2H).4p & 3G* &3& 3d3.(2G).4d & 3G &3&292513.1&378831.6&1158.5000&&0.0030&&&&1.0e-02$^\dagger$&1.29E+10$^\mathsection$&   \\
7& 3d3.(2G).4p & 3F* &4& 3d3.(4F).4d & 3F &4&280366.9&366564.3&1160.1280&&0.0040&&&&1.0e-02&1.12E+10$^\mathsection$&   \\
\end{longtable}

\normalsize

\subsection{Observed lines}

\scriptsize
\setlength\tabcolsep{3pt}
\def\arraystretch{1.2}
\begin{longtable}{*{5}{r}*{8}{c}}
\caption{\small{Observed line wavelengths in the STIS spectrum of G191-B2B. ID refers to the number in column 1 of Table \ref{tab:allwavelengths}. $\lambda_\text{obs}$ is the mean observed-frame wavelength of the absorption feature with uncertainty $\sigma(\lambda_\text{obs})$. EW is the equivalent width with uncertainty $\sigma$(EW). The absence of $\sigma$(EW) indicates that another line is nearby such that the detection algorithm interpreted the whole feature as a blend. Wavelength and equivalent widths are in {\AA}. Columns 6-11 give redshift z, b-parameter (in km/s) and column density logN (N is in atoms/cm) with their associated uncertainties. A tick in the last two columns indicates that a line was used for measuring $\alpha$.}}
\label{tab:stiswav} \\

\hline
\multicolumn{1}{c }{\textbf{ID}} &
\multicolumn{1}{c }{\textbf{$\bm{\lambda_\text{obs}}$}} &
\multicolumn{1}{c }{\textbf{$\bm{\sigma(\lambda_\text{obs})}$}} &
\multicolumn{1}{c }{\textbf{EW}} &
\multicolumn{1}{c }{\textbf{$\bm{\sigma}$(EW)}} &
\multicolumn{1}{c }{\textbf{z [$\bm{\times10^{-5}}$]}} &
\multicolumn{1}{c }{\textbf{$\bm{\sigma}$(z) [$\bm{\times10^{-5}}$]}} &
\multicolumn{1}{c }{\textbf{b}} &
\multicolumn{1}{c }{\textbf{$\bm{\sigma}$(b)}} &
\multicolumn{1}{c }{\textbf{logN}} &
\multicolumn{1}{c }{\textbf{$\bm{\sigma}$(logN)}} &
\multicolumn{1}{c}{\textbf{K14}} &
\multicolumn{1}{c}{\textbf{W19}} \\
\hline
\endfirsthead

\caption* {Table \ref{tab:stiswav} Continued: }\\
\hline
\multicolumn{1}{c }{\textbf{ID}} &
\multicolumn{1}{c }{\textbf{$\bm{\lambda_\text{obs}}$}} &
\multicolumn{1}{c }{\textbf{$\bm{\sigma(\lambda_\text{obs})}$}} &
\multicolumn{1}{c }{\textbf{EW}} &
\multicolumn{1}{c }{\textbf{$\bm{\sigma}$(EW)}} &
\multicolumn{1}{c }{\textbf{z [$\bm{\times10^{-5}}$]}} &
\multicolumn{1}{c }{\textbf{$\bm{\sigma}$(z) [$\bm{\times10^{-5}}$]}} &
\multicolumn{1}{c }{\textbf{b}} &
\multicolumn{1}{c }{\textbf{$\bm{\sigma}$(b)}} &
\multicolumn{1}{c }{\textbf{logN}} &
\multicolumn{1}{c }{\textbf{$\bm{\sigma}$(logN)}} &
\multicolumn{1}{c}{\textbf{K14}} &
\multicolumn{1}{c}{\textbf{W19}} \\
\hline
\endhead
\hline
\endfoot
\hline
\endlastfoot
27&1250.833&0.001&0.003&&7.88&0.06&4.99&0.30&12.12&0.02&$\checkmark$& \\
36&1280.573&0.0012&0.00617&0.000197&8.17&0.04&5.36&0.18&13.22&0.01&$\checkmark$&$\checkmark$ \\
38&1281.468&0.001&0.002&&7.93&0.11&5.11&0.59&12.15&0.04&$\checkmark$& \\
43&1286.015&0.001&0.002&&7.73&0.06&3.68&0.30&12.51&0.02&$\checkmark$&$\checkmark$ \\
45&1288.269&0.0009&0.00355&0.000148&7.89&0.04&3.98&0.20&12.73&0.01&$\checkmark$&$\checkmark$ \\
48&1291.298&0.002&0.005&&8.24&0.07&6.50&0.30&12.90&0.02&$\checkmark$&$\checkmark$ \\
50&1293.409&0.001&0.003&&7.85&0.06&5.47&0.27&13.09&0.02&$\checkmark$& \\
51&1293.485&0.001&0.003&&7.92&0.04&4.40&0.18&12.85&0.01&$\checkmark$& \\
\end{longtable}

\normalsize

\bsp
\label{lastpage}
\end{document}